\documentclass[prd,aps,preprint,showpacs]{revtex4}
\usepackage{graphicx}
\usepackage{dcolumn}
\usepackage{amsfonts}

\def \nn{\nonumber}

\newcommand{\ep}{\epsilon}

\newcommand{\vphi}{\varphi}

\newcommand{\pa}{\partial}
\newcommand{\td}{\tilde}

\newcommand{\beq}[1]{\begin{eqnarray}\label{#1}}
\newcommand{\eeq}{\end{eqnarray}}

\newcommand{\CN}[1]{{\cal N}=#1}

\begin{document}
\title{Dynamics of Giant-Gravitons in the LLM geometry  \\
and the Fractional Quantum Hall Effect}
\author{Jian Dai$^{1}$}\email{jdai@physics.utah.edu}
\author{Xiao-Jun Wang$^2$} \email{wangxj@ustc.edu.cn}
\author{Yong-Shi Wu$^{1,3}$}\email{wu@physics.utah.edu}
\affiliation{\centerline{$^1$ Department of Physics, University of
Utah} \centerline{Salt Lake City, Utah 84112, USA}
\centerline{$^2$ Interdisciplinary Center for Theoretical Study}
\centerline{University of Science and Technology of China}
\centerline{AnHui, HeFei 230026, China}
\centerline{$^3$ Institute for Solid State Physics,
University of Tokyo} \centerline{Kashiwanoha 5-1-5, Kashiwa, Chiba
277-8581, Japan}}
\begin{abstract}
The LLM's 1/2 BPS solutions of IIB supergravity are known to be
closely related to the integer quantum Hall droplets with filling
factor $\nu=1$, and the giant gravitons in the LLM geometry behave
like the quasi-holes in those droplets. In this paper we consider
how the fractional quantum Hall effect may arise in this context,
by studying the dynamics of giant graviton probes in a special LLM
geometry, the $AdS_5\times S^5$ background, that corresponds to a
circular droplet. The giant gravitons we study are D3-branes
wrapping on a 3-sphere in $S^5$. Their low energy world-volume
theory, truncated to the 1/2 BPS sector, is shown to be described
by a Chern-Simons finite-matrix model. We demonstrate that these
giant gravitons may condense at right density further into
fractional quantum Hall fluid due to the repulsive interaction in
the model, giving rise to the new states in IIB string theory.
Some features of the novel physics of these new states are
discussed.
\end{abstract}
\preprint{USTC-ICTS-05-09} \maketitle

\section{Introduction}

The Quantum Hall Effect (QHE), a novel phenomenon of 2d electron
gas in a strong transversal magnetic field, has been shown to have
intriguing relationship with string/M-theory. In the early days,
the realization of the quantum Hall systems in terms of solitonic
systems in string/M theory (M-branes or D-branes plus strings
etc.) is the major focus \cite{BBST00,GR00,Susskind01a}. Recently
this issue has attracted new interests in the context of AdS/CFT
correspondence due to the progress in understanding of the 1/2 BPS
sector on both string theory and gauge theory sides.

In a remarkable work \cite{LLM}, Lin, Lunin and Maldacena (LLM)
found a general class of static, non-singular solutions of IIB
supergravity; these solutions preserve at least 16 supercharges
together with a $SO(4)\times SO(4)\times R$ global symmetry, where
$R$ is the time translation symmetry. The maximally supersymmetric
solution $AdS_5\times S^5$ is a special case of the LLM solutions.
For an observer at infinity, the solutions have total energy equal
to angular momentum, which is just the 1/2 BPS condition in
$\CN{4}$ supersymmetric Yang-Mills theory (SYM) in 4 dimensions.
Therefore, the LLM geometries as semi-classical gravitational
excitations in Type IIB string theory on $AdS_5\times S^5$
background correspond to the 1/2 BPS sector in the dual CFT, i.e.
the $\CN{4}$ SYM in 4d.

LLM's interest in finding the gravitational dual of the 1/2 BPS
sector in $\CN{4}$ SYM was inspired by \cite{CJR01,Ber04a}. In
\cite{Ber04a}, Berenstein made the following observation: certain
limiting procedure produces a decoupled 1/2 BPS sector in AdS/CFT
correspondence and the self consistency of this decoupling
procedure roots in the fact that all other degrees of freedom cost
too much energy and can be integrated out. Accordingly, the study
of the 1/2 BPS sector as LLM conducted provides a new testing

ground for AdS/CFT correspondence, which is more controllable on
both sides than the entire theories. Also it was noted in
\cite{Ber04a} that the 1/2 BPS sector in $\CN{4}$ SYM can be
mapped to a system of 1d free fermions, which in turn can be
mapped to quantum Hall droplets with filling factor $\nu=1$ in 2d
phase space. (This observation was further substantiated in refs.
\cite{Ber04b,GMSS05,MR05,TT05}.)

LLM geometries provide us with an ideal setting for studying the
AdS/QH/SYM connection on the string theory side. This is because
the LLM solutions are completely determined by the boundary value
of a single function $z$ of three non-compact spatial coordinates
$(x_1,x_2,y)$ with $y\geq 0$. When the boundary value of $z$ on
the two dimensional plane at $y=0$ is taken to be $\pm 1/2$, the
corresponding geometries are non-singular. This boundary value can
be interpreted as the distributions for two types of point charge
sources on the two dimensional boundary plane. R-R five-flux
quantization infers that the areas of connected regions for $z=\pm
1/2$ on the boundary plane to be quantized in appropriate unit.
Therefore, one may view the either regions with $z=1/2$ or
$z=-1/2$ on $y=0$ plane as droplets of an incompressible fluid,
like the quantum Hall fluid. The special case $AdS_5\times S^5$
corresponds a uniform distribution of point sources in a circular
disk with $z=-1/2$ on the $y=0$ plane so it corresponds to a
circular quantum Hall droplet on the plane. In this way, it is
convenient to describe the $AdS_5\times S^5$ geometry and its
small deformations (the so-called bubbling AdS geometries) in
terms of quantum Hall droplets \cite{LLM,CS04}.

Since Integer Quantum Hall Effect (IQHE) ($\nu=1$ for filling
fraction) emerges naturally in the LLM geometries, one can't help
but wonder whether this is merely an accidental analogy or this
signals any profound physics? More precisely, one may wonder
whether the knowledge of QHE in condensed matter physics can
provide any insights into the dynamics of the LLM geometries. In
this paper, we make progress in this direction by examining
whether {\it Fractional Quantum Hall Effect} (FQHE) can make its
way into the dynamics of the LLM geometries. In condensed matter
physics, it is well-known that the quasi-particles or quasi-holes
in the $\nu=1$ quantum Hall system can condense to form new
quantum Hall liquid states as an effect of interactions. For
example, the simplest case of $\nu=2/3$ FQH state can be viewed as
the condensation of the quasi-holes in the $\nu=1$ IQH fluid.

As we just mentioned, $AdS_5\times S^5$ background is treated as a
circular QH droplet in the LLM/IQH analogy. Adding a giant
graviton in $S^5$ which is a D3-brane wrapping on a 3-sphere in
$S^5$ \cite{MST00} corresponds to a quasi-hole excitation
\cite{LLM} while adding a giant graviton in $AdS_5$
\cite{Myers00,HHI00} corresponds to a quasi-particle excitation.
For simplicity and definitude, we will only examine the dynamics
of quasi-hole type of giant gravitons, and treat these giant
gravitons as probes to the $AdS_5\times S^5$ background (as a
special LLM geometry). Recall that the five-form flux of
$AdS_5\times S^5$ background plays the role of the constant
magnetic field on the $y=0$ plane. We will show that the
low-energy effective action for the giant gravitons of quasi-hole
type, when truncated by 1/2 BPS condition, exhibits all of the
essential features of a QH system. In other words, the giant
gravitons in the 1/2 BPS sector behave just like the charged
particles in a strong magnetic field.

For one single giant graviton, the dynamics of the Landau problem
for a single quasi-hole is exactly reproduces. This result is just
a self-consistent check that the $AdS_5\times S^5$ background

behaves as an unexcited $\nu=1$ QH droplet. For many giant
gravitons, if their number is much smaller than the background
flux, we can ignore their back-reaction. Then the non-abelian
low-energy dynamics (world-volume theory) of the giant graviton
probes leads to a Chern-Simons matrix model; this matrix model is
exactly the matrix description of the fluid dynamics of a system
of non-relativistic charged particles moving in a constant
magnetic field, i.e. the infinite-matrix model in ref.
\cite{Susskind01a} for the bulk QH state, or the finite-matrix
model in ref. \cite{Poly01} that incorporates edge excitations.
Here the matrix description for the QH fluid is remarkably
suitable for our purpose in the context of giant graviton
dynamics: The expectation values of the diagonal elements are the
positions of giant gravitons in the transverse space, while the
off-diagonal elements describe open strings stretching between
different giant gravitons, giving rise to the interaction between
giant gravitons (as quasi-holes). Ignoring the off-diagonal
elements and following the arguments in ref. \cite{Ber04a}, we
will show that this matrix model in the eigenvalue basis reduces
to a free fermion system, provided all of the eigenvalues are
non-degenerate. It implies that the world-volume gauge symmetry of
$M$ giant gravitons is broken from $U(M)$ down to $U(1)^M$, and
each giant graviton fills a state with different angular momentum
$j$.
Including off-diagonal elements will lead to {\it repulsive}
interactions between giant gravitons, in favor of forming new
incompressible fluids with fractional filling factors (FQH fluid).
This is consistent with the fact that the level of the resulting
Chern-Simons matrix model takes only integer values, whose
inverses correspond to the fractional filling factors. In this
manner, we are able to demonstrate that the 1/2 BPS dynamics of
the giant gravitons wrapping on the three-sphere in $S^5$, which

behave like quasi-holes in an IQH droplet with $\nu=1$, allows the
formation of a ground state corresponding to a FQH liquid due to
quasi-hole condensation. Moreover, when the number $M$ of the
giant gravitons is finite but large, the resulting finite-matrix
model allows gapless edge excitations, in a manner similar to that
pointed out by Polychronakos \cite{Poly01}. Some of the novel
properties of the new states can be understood by referring to the
knowledge of the FQH fluid in condensed matter physics. We
speculate that, upon including back-reactions of the giant
gravitons, the FQH state would accordingly give rise to new
geometries with singularities in IIB string theory, and that the
novel properties associated with the FQH state may help us
understand how quantum effects resolve the singularities.

This paper is organized as follows. In Section II we briefly
review the LLM's 1/2 BPS solutions in IIB supergravity, in
particular the $AdS_5\times S^5$ geometry from LLM solutions. We
also review giant gravitons in $AdS_5\times S^5$. Part of the
purpose of this review is to set up the notations. In Section III
we derive the effective theory which describes low energy dynamics
in a special LLM geometry, the $AdS_5\times S^5$ background, of
giant gravitons moving in $S^5$. We shall show how this model
reproduces the Landau problem for a single giant graviton. Then in
Section IV, it is shown how to pass from the effective field
theory for giant gravitons to a Chern-Simon finite-matrix model,
which turns out to be essentially the same model proposed in ref.
\cite{Poly01}. In Section V we shall demonstrate that the
resulting matrix model for giant gravitons accommodates all
essential features of FQHE, including the repulsive interaction

that favors the formation of the FQH state. And we speculate that
novel properties of the FQH state may help us understand how
quantum effects resolve the singularities in the new geometry that
emerges when back-reactions are included. Finally, Section VI is
devoted to a brief summary.

\section{1/2 BPS geometries in IIB theory and giant graviton probes}

The class of 1/2 BPS states is known to play an important role in
testing AdS/CFT correspondence. On the CFT side these states are
associated to chiral primary operators with conformal weight
$\Delta=J$, where $J$ is a certain $U(1)$ charge in $R$-symmetry
group. For small excitation energy $J\ll N$ ($N$ being the rank of
the gauge group), these BPS states on the AdS side correspond to
graviton modes propagating in the bulk. When the excitation energy
increases to $J\sim N$, some of these states are described by
giant gravitons, i.e. spherical D3-branes either in the internal
sphere\cite{MST00} or in AdS\cite{Myers00,HHI00}. As excitation
energy increases further the back-reaction cannot be ignored, and
new geometries which preserve 16 supercharges are expected to
emerge in IIB supergravity.

\subsection{LLM's IIB solutions}

In a seminal paper\cite{LLM}, LLM explicitly derived the most
general, smooth 1/2 BPS solutions of IIB supergravity, that are
invariant under $SO(4)\times SO(4)\times R$ global symmetry. They
contain the non-vanishing R-R 5-form flux of the form \beq{1}
F_{(5)}=F_{\mu\nu}dx^\mu\wedge dx^\nu\wedge d\Omega_3 +
\td{F}_{\mu\nu}dx^\mu\wedge dx^\nu\wedge d\td{\Omega}_3, \eeq
where $\mu,\nu=0,1,2,3$, and $d\Omega_3$ and $d\td{\Omega}_3$
denote the volume forms of two three-spheres. The two $SO(4)$s are
the rotational symmetries of these two three-spheres respectively.
The dilaton-axion moduli is constant and the 3-form field
strengths are set to be zero. Accordingly, LLM geometry is
determined explicitly by the following metric and R-R 5-form
measured in the unit $l_s=1$,
\beq{2}
ds^2&=&-h^{-2}(dt+V_idx^i)^2+h^2(dy^2+dx^idx^i)+ye^{G}d\Omega_3^2
+ye^{-G}d\td{\Omega}_3^2, \nn \\
h^{-2}&=&2y\cosh{G},\hspace{1in}z=\frac{1}{2}\tanh{G},  \\
y\pa_yV_i&=&\ep_{ij}\pa_jz,\hspace{1in}
y(\pa_iV_j-\pa_jV_i)=\ep_{ij}\pa_yz, \nn \\
F_{(2)}&=& F_{\mu\nu}dx^\mu\wedge dx^\nu =
dB_t\wedge dt+dA_{(1)}=d(B_tdt+B_tV+\hat{B}), \nn \\
\td{F}_{(2)}&=& \td{F}_{\mu\nu}dx^\mu\wedge dx^\nu=
d\td{B}_t\wedge dt+d\td{A}_{(1)}
=d(\td{B}_tdt+\td{B}_tV+\hat{\td{B}}), \nn \\
B_t&=&-\frac{1}{4}y^2e^{2G},\hspace{1in}
\td{B}_t=-\frac{1}{4}y^2e^{-2G}, \nn \\
d\hat{B}&=&-\frac{1}{4}y^3\ast_3d\frac{z+1/2}{y^2}\hspace{0.6in}
 d\hat{\td{B}}=-\frac{1}{4}y^3\ast_3d\frac{z-1/2}{y^2},
\nn \eeq where $i=1,2$, $t=x^0$, $y=x^3$. $\ast_3$ is the flat
space epsilon symbol in the three dimensions parameterized by
$x_1,x_2,y$. The full solution is determined by a single function
$z(x_1,x_2,y)$, which obeys the Laplace equation in six dimensions
\beq{3}\pa_i\pa_i \frac{z}{y^2}+\frac{1}{y^3}\pa_y
(y^3\pa_y\frac{z}{y^2})=0, \eeq with $y$ acting like the radial
coordinate in four dimensions. This solution is non-singular as
long as $z=\pm\frac{1}{2}$ on the two dimensional plane spanned by
$(x_1,x_2)$ at $y=0$.

As long as $y > 0$, there are two 3-spheres $S^3$ and $\td{S}^3$,
corresponding to the two $SO(4)$ isometries. At the $y=0$ plane
$S^3$ shrinks to a point in the region with $z=-\frac{1}{2}$,
while $\td{S}^3$ shrinks in the region with $z=\frac{1}{2}$.
Following the setup in Section 2.4 of ref.\cite{LLM}, we may
choose a surface $\td{\Sigma}_2$ in the $(y,x_1,x_2)$ space that
ends at $y=0$ on a closed non-intersecting curve lying in a region
with $z=\frac{1}{2}$. A smooth five-manifold $\td{\Sigma}_5$ can
be constructed by the fibration of $\td{S}^3$ over
$\td{\Sigma}_2$. The five-form flux measured on this five-manifold
is given by
\beq{4}\td{N}=-\mu_3\int_{\td{\Sigma}_5} F_{(5)}=\frac{({\rm
Area})_{z=-1/2}}{2\pi}, \eeq where here the normalization
$\mu_3=T_3=(2\pi)^{-3}$ in string units is used and $({\rm
Area})_{z=-1/2}$ is the area of the region with $z=-1/2$ inside
$\td{\Sigma}$ on $y=0$ plane. We can as well construct another
five-manifold $\Sigma_5$ by the fibration of $S^3$ over a surface
$\Sigma_2$; $\Sigma_2$ ends in the region with $z=-\frac{1}{2}$ of
the $y=0$ plane. The five-form flux measured on $\Sigma_5$ is
\beq{5} N=\mu_3\int_{\Sigma_5} F_{(5)} =\frac{({\rm
Area})_{z=1/2}}{2\pi}. \eeq Equations (\ref{4}) and (\ref{5})
indicate that the area of each connected region with either
$z=\frac{1}{2}$ or $z=-\frac{1}{2}$ on the $y=0$ plane is
quantized. This is because the total 5-flux through any
five-sphere is quantized by Dirac quantization condition.

The LLM solutions (\ref{1}) and (\ref{2}) have been shown to have
the energy equal to the angular momentum and to preserve 16
supersymmetries, so they are indeed geometries satisfying the 1/2
BPS condition. They provide us a new theoretical setting to test
the AdS/CFT duality at a new level.

\subsection{$AdS_5\times S^5$ from LLM}

The LLM geometries are completely determined by the distribution
of the sources for the function $z$ on the boundary $y=0$ plane.
$AdS_5\times S^5$ as the maximal BPS geometry is a special case of
the LLM geometries, with the sources uniformly distributed in a
disk with $z=-1/2$ on the $y=0$ plane. Then the solution of the
six dimensional Laplace equation (\ref{3}) is given by \beq{6}
z(r,y;r_0)&=&\frac{r^2-r_0^2+y^2}
{2\sqrt{(r^2+r_0^2+y^2)^2-4r^2r_0^2}}, \nn \\
V_\phi(r,y;r_0)&=&\frac{1}{2}-\frac{r^2+r_0^2+y^2}
{2\sqrt{(r^2+r_0^2+y^2)^2-4r^2r_0^2}},
\eeq
where $(r,\phi)$ are the polar coordinates on the $(x_1,x_2)$
plane, and $r_0$ is the radius of the disk. For this isotropic
solution, $\hat{B}$ and $\hat{\td{B}}$ defined in~(\ref{2}) can be
expressed as
\beq{7}
\hat{B}=K(r,y;r_0)d\phi,\hspace{1in}
\hat{\td{B}}=\td{K}(r,y;r_0)d\phi,
\eeq
where $K$ and $\td{K}$ are determined by the following
differential equations
\beq{8}\pa_r
K&=&-\frac{ry}{4}\pa_y z+\frac{r}{2}(z+\frac{1}{2}),
\hspace{0.7in} \pa_yK=\frac{ry}{4}\pa_r z, \nn \\
\pa_r \td{K}&=&-\frac{ry}{4}\pa_y z+\frac{r}{2}(z-\frac{1}{2}),
\hspace{0.7in} \pa_y\td{K}=\frac{ry}{4}\pa_r z.
\eeq
It is easy to check that the above equations are compatible with
the Laplace equation~(\ref{3}).

Inserting (\ref{6}) in the LLM solution~(\ref{2}) and performing
the change of coordinate
\beq{9}y&=&r_0\sinh{\rho}\sin{\theta}, \nn \\
r&=&r_0\cosh{\rho}\cos{\theta}, \\
\phi&=&\td{\phi}-t, \nn \eeq we obtain the standard $AdS_5\times
S^5$ metric in the global coordinates \beq{10}
ds^2=r_0[-\cosh^2{\rho}dt^2+d\rho^2 +\sinh^2{\rho}d\Omega_3^2
+d\theta^2+\cos^2{\theta}d\td{\phi}^2
+\sin^2{\theta}d\td{\Omega}_3^2]. \eeq Matching it with
$AdS_5\times S^5$ geometry from the near-horizon limit of
D3-branes, we have $r_0=R^2_{AdS}=R^2_{S}=\sqrt{4\pi g_s\td{N}}$.
Meanwhile, from eq. (\ref{9}) we see that the interior of the disk
on the $y=0$ plane corresponds to $\rho=0$, i.e., the center of
$AdS_5$, while the exterior of the disk on the $y=0$ plane
corresponds to $\theta=0$, i.e., the north-pole of $S^5$.

The last equation in (\ref{9}) is linear in time. This implies
that the LLM coordinates and the global coordinates of
$AdS_5\times S^5$ differ from each other by a relative frame
rotation in the $(x_1,x_2)$ plane. In the LLM frame, the
$AdS_5\times S^5$ metric (\ref{10}) is seen by an observer who
co-moves with a rotating frame in the $(x_1,x_2)$ plane.
Conversely, in the global frame for $AdS_5\times S^5$, an observer
who moves along the $\td{\phi}$-circle in $S^5$ will see the LLM
metric (\ref{6}) determined by a disk source. So the double
scaling limit of large angular momentum and large radius of
$\td{\phi}$-circle is just the PP-wave limit of $AdS_5\times S^5$
\cite{BMN02}.

For convenience for the following discussions, we will evaluate
the $y\to 0$ limit of various functions. For the interior of the
disk, ($r^2<r_0^2$), we have
\beq{11}&&z\to -1/2, \hspace{2in}V_\phi\to\frac{r^2}{r^2-r_0^2}, \\
&&\left\{\begin{array}{l}B_t\to 0, \\
\td{B}_t\to -\frac{(r^2-r_0^2)^2}{4r_0^2}, \end{array} \right.
\hspace{1in}
\left\{\begin{array}{l}K\to 0, \nn \\
\td{K}\to -r^2/4 \end{array} \right. \eeq while for the exterior
of disk ($r^2>r_0^2$),
\beq{12}&&z\to 1/2, \hspace{2in}V_\phi\to -\frac{r_0^2}{r^2-r_0^2}, \\
&&\left\{\begin{array}{l}
B_t\to -\frac{(r^2-r_0^2)^2}{4r_0^2}, \nn \\
\td{B}_t\to 0, \end{array} \right. \hspace{1in}
\left\{\begin{array}{l}K\to r^2/4, \\
\td{K}\to 0. \end{array} \right. \eeq We see that the radii of the
two three-spheres, $S^3$ inside $AdS_5$ and $\td{S}^3$ inside
$S^5$, are proportional to $\sqrt{|B_t|}$ and $\sqrt{|\td{B}_t|}$,
respectively. Thus in the interior (exterior) of the disk at
$y=0$, $S^3$ ($\td{S}^3$) shrinks to zero size.

\subsection{Giant graviton revisited}

In this subsection, we will revisit the giant graviton in
$AdS_5\times S^5$\cite{MST00}, but use the LLM's expressions in
which the $R-R$ field strength is slightly different from that
given in refs. \cite{MST00,Myers00}. This re-formulation will be
helpful for the discussions in section V.

A giant graviton is a spherical D3-brane, which has the same
quantum numbers as a graviton. So the D3-brane wraps a 3-sphere
either in $AdS_5$ or in $S^5$. Its dynamics is governed by the
world-volume action \beq{13} S=-T_3\int d^4\xi \sqrt{-{\rm
det}(P[G]_{ab}})+T_3\int P[C_{(4)}]. \eeq where $G_{\mu\nu}$ and
$C_{(n)}$ are the background metric and R-R $n$-form potential,
respectively. $P[\cdots]$ denotes the pullback of the enclosed
spacetime tensor to the brane world-volume. For our purpose, we
only consider D3-branes wrapping on a 3-sphere $\td{S}^3$ in
$S^5$. Then we may take the static gauge: \beq{14}\xi^0=t,\;\xi^m=
\omega^m\qquad (m=1,2,3), \eeq with $\omega^m$ being the
coordinates of $\td{S}^3$, and consider a trial solution of the
form, in the polar coordinates on the $(x_1,x_2)$ plane,
\beq{15}r=\td{r}{r_0} \hspace{0.5in}(0< \td{r}={\rm const.}\leq
1), \hspace{0.5in} y=0, \hspace{0.5in} \phi=\td{\phi}(t)-t. \eeq
Substituting the background metric~(\ref{10}), the R-R
potential~(\ref{1}) and the ansatz~(\ref{15}) into the
world-volume action~(\ref{13}) and integrating over the angular
coordinates yield the following Lagrangian \beq{16}
L(t)/r_0^2=-(1-\td{r}^2)^{3/2}\sqrt{1-\td{r}^2
\dot{\td{\phi}}^2}+\td{r}^4\dot{\td{\phi}} +(1-2\td{r}^2). \eeq
The corresponding Hamiltonian and the angular momentum $J$
conjugate to $\td{\phi}$ read \beq{17}
H/r_0^2&=&\sqrt{(1-\td{r}^2)^3+\frac{1}{\td{r}^2} (j-\td{r}^4)^2}
-(1-2\td{r}^2), \nn \\
j&=&\frac{J}{r_0^2}=\frac{(1-\td{r}^2)^{3/2}\td{r}^2
\dot{\td{\phi}}} {\sqrt{1-\td{r}^2\dot{\td{\phi}}^2}}-\td{r}^4.
\eeq The above Hamiltonian possesses two local minima. One of them
is at $\td{r}=1$, the boundary of the disk, where the D3-brane
shrinks to point and behaves like an ordinary graviton. Another
minimum locates at $\td{r}=\sqrt{j}$ for fixed $j\in [0,1)$.
D3-branes at this minimum preserve non-vanishing size, and are
called ``giant gravitons'' in the literature. The energy of those
classical stable ``giant gravitons'' is given by
\beq{17a}E_g=J=\td{r}^2r_0^2=r^2. \eeq This is precisely the $1/2$
BPS condition.

It is interesting to note that one always has $\dot{\td{\phi}}=1$
at these minima, independent of $J$. For an observer co-moving
with the giant graviton, he/she sees that the world-volume
dynamics along the circle is labeled by the variable
$\vphi=\td{\phi}-t$. In other words, the Lagrangian~(\ref{16})
should be rewritten as
\beq{18}
L(t)=\frac{1}{2}r^2\dot{\vphi}^2 +r^2\dot{\vphi}+\cdots.
\eeq
It is obvious that a non-zero $\dot{\vphi}$ implies quantum
fluctuations of a giant graviton around its classical stationary
points.

\section{1/2 BPS Dynamics of Giant Gravitons in LLM Geometry}

Now let us add a few giant gravitons in the LLM geometry. Suppose
the number of giant gravitons, $M$, is much less than the number
of background five-form flux $\td{N}$, so that the back-reaction
of the giant gravitons on background geometry can be ignored. They
are assumed to be located at $\rho=0$ (hence inside the $z=-1/2$
disk on the $y=0$ plane) and to wrap $\td{S}^3$ in $S^5$. We shall
show that the background R-R flux provides a homogeneous magnetic
field perpendicular to the $y=0$ plane and the dynamics of giant
graviton in the $(x_1,x_2)$ plane at $y=0$ is the same as that of
electrons moving in a 2d plane with a constant transversal
magnetic field.

The low energy dynamics of giant gravitons in a fixed background
(without NS $B$-field) is known \cite{Myers99} to be described by
the following non-Abelian generalization  of the DBI action
(\ref{13}) \beq{19} S_{DBI}=-T_3\int d^{4}\xi\;{\rm STr}
\sqrt{-{\rm det}(P[G_{ab}+G_{ai}(Q^{-1}-\delta)^{ij}G_{jb}] +2\pi
F_{ab}) {\rm det}(Q^i_{\ j})}, \eeq with$^1$ \footnotetext[1]{We
use $i,j,k$ to denote the directions transverse to the branes, and
$a,b,c$ those along the branes.} \beq{20} Q^i_{\ j}&=&\delta^i_{\
j}+i[X^i,X^k]G_{kj}, \eeq plus the Chern-Simons action
\cite{LiM95}, with the bosonic part \beq{21}S_{CS}=T_3\int {\rm
STr}\left(P[e^{i {\rm i}_{_X}{\rm i}_{_X}}(\sum_n C^{(n)})]e^{2\pi
F}\right), \eeq supplemented by proper fermionic terms . Here the
transverse coordinates $X^i$, as world-volume scalar fields, are
$M\times M$ matrices in the adjoint representation of the $U(M)$
gauge group. STr($\cdots$) denotes a symmetrical trace in gauge

group. The operator i$_{_X}$ is the contraction with $X_i$:
\beq{22} {\rm i}_{_X}{\rm i}_{_X}C^{(n)}&=&\frac{1}{(n-2)!}X^{i_1}
X^{i_2} C^{(n)}_{i_1i_2i_3\cdots i_n}dx^{i_3}\wedge\cdots\wedge
dx^{i_n}.
\eeq

Now we consider the giant gravitons locating at the center of AdS
($\rho=0\;\Rightarrow\;y=0$) and, as D3-branes, all wrapping on
the same $\td{S}^3$ in $S^5$. We take the static gauge (\ref{14})
again for the world-volume coordinates. Recall that our purpose is
to study the 1/2 BPS sector in AdS/CFT duality, treated as a
closed sector in view of the arguments given in ref.
\cite{Ber04a}. Thus we are allowed to truncate the degrees of
freedom of the giant gravitons to those in the 1/2 BPS sector:
Namely the world-volume gauge field is taken to be zero, and we
turn on the transverse scalar fields along $x_1,x_2$ directions
only. (According to the arguments in ref. \cite{Ber04a}, there
exists a limit in which all other degrees of freedom are
decoupled, i.e. they cost too much energy and therefore can be
integrated out at low energy.) Thus by inserting the background
metric and five-form field strength given by (\ref{1}), (\ref{2})
and (\ref{6}) into (\ref{19}) and (\ref{21}), we obtain \beq{23}
S_{BI}&=&-2\pi^2 T_3\int dt\; {\rm Tr}\{4|\td{B}_t|
+4|\td{B}_t|V_i \dot{X}_{i}-\frac{1}{2} \dot{X}_{i}^2
+\frac{1}{2}[X_1,X_2]^2\}+\cdots, \eeq and \beq{24}
S_{CS}&=&-8\pi^2 T_3\int dt \;{\rm Tr}\{\td{B}_t+(\td{B}_tV_i
+\hat{\td{B}}_i) \dot{X}_{i}\}+\cdots. \eeq Here $X_{i}$'s have
been restricted to the lowest order modes in the expansion of
spheric harmonics on $\td{S^3}$, because we focus on the 1/2 BPS
states. The dots ``$\cdots$'' denotes higher order terms in
powers of $M/N$ that are suppressed in our probe approximation.

The quartic term plays no role in the 1/2 BPS sector. Using
eqs.~(\ref{11}) and (\ref{12}) in (\ref{23}) and (\ref{24}), we
get a matrix model with the following Lagrangian:
\beq{25}
L(t)=\frac{1}{2l_s}{\rm Tr}\dot{X}_i^2+\frac{1}{l_s^2}
\ep_{ij}{\rm Tr}X^i\dot{X}^j,
\eeq
where we have restored the $l_s$-dependence and rescaled $X_i$ by
$X_i\to \sqrt{4\pi} X_i$.

For a single giant graviton, with $M=1$, $X_1$ and $X_2$ become
real numbers; then in polar coordinates and with $\rho=0$, the
above Lagrangian is reduced to eq. (\ref{18}). This consistency
implies that the Lagrangian~(\ref{25}) describes the quantum
fluctuations of the giant graviton around their classical
stationary point in the 1/2 BPS sector. On the other hand, the
Lagrangian~(\ref{25}) with $M=1$ precisely describes the Landau
problem, for a non-relativistic charged particle with mass
$m=1/l_s$ moving in a constant magnetic field $B=2/l_s^2$.

For the number of giant gravitons $M>1$, $X_i$'s are $M\times M$
matrices. Their diagonal elements describe the positions of the
giant gravitons, while the off-diagonal elements describe the
interactions between the giant gravitons. At low energy, one may
take the limit $l_s\to 0$. This is equivalent to take the limit
$B/m\to\infty$, i.e. the cyclotron frequency in the Landau problem
tends to infinity. However, for a large number of giant gravitons
the classical Lagrangian~(\ref{25}) does not provide a proper
description for the quantum giant-graviton fluid, since the
2-plane should become a non-commutative 2-plane when all giant
gravitons are projected down to the lowest Landau levels (LLL). In
the next section, we shall show that the matrix model (\ref{25})
has to be supplemented by a non-commutativity constraint in an
incompressible fluid phase, so that the quantum behavior of the
many giant graviton system is that of a quantum Hall fluid,
described by a Chern-Simons matrix model.

To conclude this section, we note that the Lagrangian (\ref{25})
can be recast into a manifestly $U(M)$ gauge invariant form:
\beq{26}L(t)=\frac{m}{2}{\rm
Tr}(DX_i)^2+\frac{B}{2} \ep_{ij}{\rm Tr}(X^iDX^j),
\eeq
where $DX_i=\dot{X}_i+i[A_0,X_i]$ with $A_0$ the electric
potential.

\section{Chern-Simons finite-matrix model for giant gravitons}

Since the essential non-commutativity condition is absent, the
matrix model~(\ref{26}) is not sufficient to describe a QH fluid.
We shall show that this condition naturally emerges as soon as the
quantum physics of giant gravitons is considered.

\subsection{Symmetry breaking}

Since the Lagrangian~(\ref{26}) is invariant under $U(M)$ gauge
symmetry, one of the $X$s, e.g. $X_1$, can be diagonalized by a
gauge transformation$^2$. \footnotetext[2]{Here we do not adopt
the argument of using a {\it complexified} gauge group to
diagonalize $X_1$ and $X_2$ simultaneously.} In this eigenvalue
basis, with notation $(X_1)_{mn}=\delta_{mn}x_{1m},
(X_2)_{mn}=y_{mn}, y_{nn}=x_{2n}$, a typical classical Lagrangian
of the matrix model reads
\beq{27}
L=\frac{m}{2}\sum_{i,n}\dot{x}_{in}^2+ \frac{m}{2}\sum_{m\neq n}
\dot{\bar{y}}_{mn}\dot{y}_{mn}+\frac{B}{2}\sum_{i,n}
\ep_{ij}x_n^i\dot{x}_n^j -U(x,y).
\eeq
However, quantum mechanically there is a non-trivial change of
measure in path integral from the matrix-element basis to the
eigenvalue basis. So the Hamiltonian in the quantum theory is
given by \beq{28}
H_q&=&-\frac{1}{2m}\sum_{i,n}\frac{1}{\Delta^2}(\frac{\pa}{\pa
x_{in}}-iB\ep_{ij}x_j)\Delta^2(\frac{\pa}{\pa
x_{in}}-iB\ep_{ij}x_j) -\frac{1}{2m}\sum_{m, n}\frac{\pa}{\pa
\bar{y}_{mn}}\frac{\pa}{\pa y_{mn}}
+U(x,y) \nn \\
&=&\frac{1}{\Delta(x_1)}\td{H}\Delta(x_1), \eeq where
$\Delta(x)=\prod_{n<m}(x_n-x_m)$ is the Van der Monde determinant,
and $\td{H}$ is the Hamiltonian corresponding to the
Lagrangian~(\ref{27}).

The eigenfunction $\td{\psi}$ of $\td{H}$ is related to an
eigenfunction $\psi$ of $H_q$ by
$\td{\psi}(x,y)=\Delta(x_1)\psi(x,y)$. It implies that the
Hamiltonian~(\ref{27}) in the eigenvalue basis describes a quantum
system of particles obeying Fermi statistics$^3$
\footnotetext[3]{This is a 1-d fermion system instead of 2-d one,
unlike that extracted from $\CN{4}$ SYM\cite{Ber04a}. It indeed
makes sense at $m/B\to 0$ because the phase space is two
dimensional at this limit.} from the requirement that any pair of
$x_{1n}$ are different. In other words, we may have two
perspectives on the physics of giant gravitons at the quantum
level. One is to use the Hamiltonian~(\ref{28}) of $U(M)$ gauge
symmetry to describe the quantum physics, with $M$ giant gravitons
understood semiclassically coinciding. Another perspective is to
just use the Hamiltonian obtained from the classical
Lagrangian~(\ref{27}) to describe the quantum fluctuations of
giant gravitons, leaving $M$ giant gravitons separated from each
other on the $(x_1,x_2)$-plane. In the latter case the
world-volume gauge symmetry is broken from $U(M)$ to $U(1)^M$. So
the degrees of freedom corresponding to the off-diagonal elements
of $X_2$, which give excitation modes of open string stretched
between different giant gravitons, become heavy and are frozen out
at the low energy limit $l_s\to 0$. In continuous fluid
description, two different perspectives were referred as two
different fluids, the string fluid and brane fluid\cite{FST01}.
They should be related to each other, like the correspondence
between matrix basis and eigenvalue basis in the toy matrix model
discussed by Berenstein\cite{Ber04a}. For our purpose, the latter
perspective is more convenient. Thus we end up with the following
Lagrangian from Eq.~(\ref{26}):
\beq{29}
L=\sum_{n}\left[\frac{m}{2}\dot{x}_{in}^2+\frac{B}{2}
\ep_{ij}x_n^i\dot{x}_n^j-V(x_{in})\right],
\eeq
supplemented with
Fermi statistics condition. Here we have included a potential
$V(x_{in})$, which is assumed to arise from short-range
interactions and will be shown in the next section to naturally
emerge when we choose a correct vacuum. The Lagrangian (\ref{29})
describes a collection of non-relativistic charged fermions moving
in a plane subject to a perpendicular magnetic field. In the
$m/B\to 0$ limit, all particles are projected into the LLL, which
is infinitely degenerated. The degenerated states in the LLL are
distinguished by the quantum number of angular momentum $j$. Due
to Fermi statistics, different giant gravitons occupy states with
different $j$.

\subsection{Incompressible Fluid description and non-commutative
Chern-Simons theory$^4$}

\footnotetext[4]{This subsection does not contain new results. It
just follows the treatment in \cite{BSD,Susskind01a}, to show how
non-commutativity comes about in our matrix model. The readers
familiar with this issue may skip this subsection.}

Following refs. \cite{BSD,Susskind01a}, we treat the long distance
behavior of the above system as a dissipationelss fluid. The
discrete label $n$ is replaced by a pair of continuous coordinates
$(y_1,y_2)$; these coordinates are co-moving coordinates subject
to the condition that the density of particle number, $\rho_0$, is
constant. Accordingly, the Lagrangian~(\ref{29}) can be rewritten
as
\beq{30} L(t)=\int d^2y\;\rho_0\left[\frac{m}{2}\dot{x}_{i}^2(y)
+\frac{B}{2} \ep_{ij}x_i(y)\dot{x}_j(y) -V(\rho_0|\frac{\pa y}{\pa
x}|)\right]. \eeq The potential $V$ plays a role when distances
among giant gravitons are small, i.e., it arises out of
short-range forces. It leads to an equilibrium when the real space
density is $\rho_0$. The fact that such an equilibrium state
reaches the minimal energy is essentially an assumption, which may
not be true if the giant gravitons are in, say, a Wigner crystal
state. In the following we shall always make this assumption as a
working hypothesis. Thus, the Jacobian $|\pa y/\pa x|$ equals to 1
in equilibrium.

The Lagrangian (\ref{30}) is invariant under the area preserving
diffeomorphisms (APD) of $y$-plane, generated by the infinitesimal
transformations \beq{31} \delta
y_i=\ep_{ij}\frac{\pa\alpha(y)}{\pa y_j},\quad \Rightarrow \quad
\delta x_i=\ep_{jk} \frac{\pa x_i}{\pa y_j}\frac{\pa\alpha(y)}{\pa
y_k}. \eeq Small deviations from the equilibrium can be
parameterized by a vector field $A_i$, defined by
\beq{32}x_i=y_i+\ep_{ij}\theta A_j, \eeq where $\theta\ll 1$ is a
convenient parameter to control the expansion. The APD~(\ref{31})
leads to the following transformation for $A$: \beq{33} \delta
A_i=\frac{\pa\td{\alpha}}{\pa y_i} +\ep_{lm}\theta\frac{\pa
A_i}{\pa y_l} \frac{\pa\td{\alpha}}{\pa y_m}, \eeq where
$\td{\alpha}=\alpha/\theta$. The first term is in the standard
form of an Abelian gauge transformation.

Moreover, the APD invariance implies that there exists a conserved
charge, given by \beq{34} \int d^2y \Pi_i\delta x_i. \eeq In the
limit $m/B\to 0$, the conjugate momentum density $\Pi_i\propto
\ep_{ij}x^j$. Then the conserved current is the Jacobian from $x$
to $y$. Thus, the equation of motion is supplemented with the
constraint \beq{35} |\frac{\pa x}{\pa
y}|=\frac{1}{2}\ep_{ij}\ep_{mn} \frac{\pa x_m}{\pa y_i}\frac{\pa
x_n}{\pa y_j}=1. \eeq Namely the fluid density in $x$-space is
constant, or the fluid is incompressible, since by definition the
fluid density is constant in $y$-coordinates. This constraint can
be viewed as a Gauss law constraint and can be added into the
action via a Lagrangian multiplier $A_0$ . Then in the limit
$m/B\to 0$, the Lagrangian reads \beq{36} L(t)=\int
d^2y\;\rho_0\left[\frac{B}{2}
\ep_{ij}(\dot{x}^i-\theta\{x_i,A_0\})x^j +\theta A_0
-V(\rho_0|\frac{\pa y}{\pa x}|)\right], \eeq where the Poisson
bracket is defined by \beq{37} \{F(y),G(y)\}=\ep_{ij}\pa_i F\pa_j
G.
\eeq

Eqs.~(\ref{33}) and (\ref{37}) exhibit the structure of
non-commutative $U(1)$ gauge theory with spatial-spatial
non-commutativity: To the first order in $\theta$ expansion, eq.
(\ref{33}) is nothing but non-commutative gauge transformation,
while eq. (\ref{37}) is the non-commutative commutator. This
observation led Susskind to propose \cite{Susskind01a} that,
beyond the linear order, the APD invariance requires the action to
be the non-commutative Chern-Simons (NCCS) theory (with
$\nu=1/B\theta$)
\beq{38} L_{NCCS}=\frac{1}{4\pi\nu}\ep^{\mu\nu\rho} \left(
A_\mu\star\pa_\nu A_\rho +\frac{2i}{3}A_\mu\star A_\nu\star
A_\rho\right), \eeq and that (\ref{38}) provides the correct
framework for the quantum Hall system with $\nu$ being the filling
fraction. The NCCS action (\ref{38}) can be obtained from a
non-commutative matrix mechanics. The NC matrix mechanics is given
by \beq{39} L=\frac{B}{2}\ep_{ij}{\rm Tr}(\dot{X}_i+i[A_0,X_i])X_j
+B\theta {\rm Tr}A_0, \eeq with $X_i=y_i+\ep_{ij}\theta A_j$. Here
the constant matrices $y_i$ are chosen to satisfy
$[y_i,y_j]=i\theta\ep_{ij}$.

Collecting the above ingredients together, we conclude that one
should add a term $B\theta {\rm Tr}A_0$ into our matrix
model~(\ref{26}) in the large $B$ limit. This term yields a
non-commutativity constraint, and compensates the anomaly arising
from the change of measure in the large $B$ limit. The matrix
Lagrangian~(\ref{26}) plus the additional term $B\theta {\rm
Tr}A_0$ will be shown to describe a quantum Hall system of giant
gravitons. In this way, we have demonstrated that the 1/2 BPS
dynamics of giant gravitons in the LLM geometry background is
essentially that of a quantum Hall system.

\subsection{Edge excitations and a finite-matrix model for QHE}

The Gauss law constraint that follows from the Lagrangian
(\ref{39}) implies the non-commutativity
\beq{40} [X_1,X_2]=i\theta,
\eeq
which can be solved only with infinite matrices. For finite
matrices of order $M$, this constraint can be satisfied only up to
order $1/M$. To obtain a finite-matrix model with the Gauss law
exactly satisfied, one needs to add extra degrees of freedom. In
our case, we argue this necessity as follows. Since different
giant gravitons occupy different states labelled by angular
momentum $j$ of the LLL, the distance between two nearest giant
gravitons is of order $\sqrt{j/B}-\sqrt{(j-1)/B}\sim
l_s/\sqrt{j}$. When $j$ is large enough, open string excitations
stretched between the nearest giant gravitons may become very
light and can be excited at low energy. Obviously, these almost
gapless excitations prefer $j$ as large as possible, namely in the
region close to the boundary of the quantum Hall droplet. In the
fluid description, we may introduce a boundary Lagrangian to
describe these excitations:
\beq{41}L_{b}=\int d^2y \rho_0\delta(\Gamma(y_1,y_2))
\left[\frac{m_b}{2}\dot{\phi}^*(y) \dot{\phi}(y)
+\frac{i}{2}B(\dot{\phi}^*\phi-\phi^*\dot{\phi})
-\frac{B\mu_b}{2}\phi^*(y)\phi(y)\right],
\eeq
where
$\Gamma(y_1,y_2)=0$ defines the boundary, $\phi(y)$ is a complex
field defined on the boundary only and $\mu_b\sim 1/(Ml_s)$. In
fact, there are apparently two boundary fields and, later, we will
show that they correspond to the continuum limit of the
off-diagonal elements $(X_i)_{1n}$ and $(X_i)_{n1}$ or their
complex combination
\begin{equation}
 \Psi_{n1}=(X_1)_{n1}+i(X_2)_{n1},~
 \Psi_{1n}=(X_1)_{1n}+i(X_2)_{1n}.
\end{equation}
(Here the convention is that the giant graviton with the largest
$j$ is labelled by index ``$1$''). However, half of them can
always be gauged away by the residual $U(1)^M$ gauge symmetry, and
only one boundary field is physical. (This corresponds to the
well-known fact that the edge states of a quantum Hall droplet are
chiral, in the sense that edge wave propagates along the boundary
only in one direction, not in the opposite direction.)

Under an infinitesimal APD, $\phi(y)$ transforms as~(\ref{31}).
But the conserved charge now is given by
\beq{42} \int
d^2y\;\rho_0\Pi_i\delta x_i +\int d^2y
\rho_0\delta(\Gamma(y_1,y_2)) (\Pi_\phi\delta\phi+
\Pi_{\phi^*}\delta\phi^*).
\eeq
For small $B$, it implies a
vortex-like excitation at the boundary. In the large-$B$ limit, it
changes the constraint (\ref{35}) to
\beq{43}
\frac{1}{2}\ep_{ab}\{x_a,x_b\}-2i\{\phi^*,\phi\}\delta(\Gamma)=1.
\eeq
Thus, with a large $B$, we get a Lagrangian to describe this
fluid with edge excitations:
\beq{44} L(t)&=&\int
d^2y\;\rho_0\left[\frac{B}{2}
\ep_{ij}(\dot{x}^i-\theta\{x_i,A_0\})x^j +\theta A_0
-V(\rho_0|\frac{\pa y}{\pa x}|)\right] \nn \\
&&-\int d^2y\;\rho_0\delta(\Gamma)
\left[iB\phi^*(\dot{\phi}-\theta\{A_0,\phi\}) +\frac{B\mu_b}{2}
\phi^*\phi\right].
\eeq
It has been mentioned in the previous
subsection that $\{A,B\}$ is the first order truncation of the
non-commutative commutator $[A,B]_\star=A\star B-B\star A$ defined
by star product, and the full expression is assumed to be the
result of the substitution of Poisson brackets by the
non-commutative commutators. Following the same logic, we pass
from the above Lagrangian to the matrix description by the
replacements \beq{45} &&\int d^2y\;\rho_0\to {\rm Tr},
\hspace{1in} \theta\{A,B\}\to [A,B], \nn
\\ &&\int d^2y\;\rho_0\delta(\Gamma) A\to A_{11},
\eeq where again the index ``$1$'' is referred to the giant graviton
with the maximal angular momentum $j$ in LLL. In particular, one
has
\beq{46} \theta\int d^2y\;\rho_0\delta(\Gamma)\phi^*\{A_0,\phi\} \to
\Psi_{1m}^\dag (A_0)_{mn}\Psi_{n1}
-\Psi_{1m}(A_0)_{mn}\Psi_{n1}^\dag. \eeq We have mentioned that
half of degrees of freedom in $\Psi_{1m}$ and $\Psi_{m1}$ may be
gauged away. Finally we obtain a matrix model described by the
following $U(M)$ invariant Lagrangian:
\beq{47} L=\frac{B}{2}{\rm
Tr}(\ep_{ij}X^iDX^j+\theta A_0) +B\Psi^\dag (iD\Psi-\mu_b\Psi),
\eeq
where $\Psi$ transforms as fundamental representation under
$U(M)$. In the above Lagrangian we have taken the limit
$m/B\to 0$.

The matrix model presented in (\ref{47}) is slightly different
from Polychronakos's finite matrix model\cite{Poly01} by the
absence of a confining $X^2$ potential. However, eq.~(\ref{17a})
indicates that the giant gravitons are actually confined around
the origin to minimalize their energy. It means that the matrix
Lagrangian~(\ref{47}), though capturing the most important
features of the giant graviton system, does not accommodate all
information of the system. To be precise, the Hamiltonian, $H'$,
obtained from the matrix Lagrangian~(\ref{47}) describes only the
quantum fluctuations of giant gravitons around the classical
stationary point or the corresponding 1/2 BPS geometries/states.
In other words, we should identify $H'=H-H_0$, where $H_0=J$ is
the energy of 1/2 BPS states. Because the angular momentum $J$
should be treated quantum mechanically, the system must be
quantized with the full Hamiltonian $H$, instead of merely $H'$.
Recalling that $H'=H_\Psi=B\mu_b\Psi^\dag \Psi$, supplemented with
the Gauss law constraint, the full Hamiltonian is of the following
form:
\beq{47d}
H=J+H_\Psi=\frac{B}{2m}{\rm Tr} (X_1\frac{\pa L}{\pa \dot{X}_2}
- X_2\frac{\pa L}{\pa \dot{X}_1})+H_\Psi
=\frac{B^2}{4m}{\rm Tr} (X_1^2+X_2^2)+H_\Psi.
\eeq
The $X$-part of the above Hamiltonian behaves as $M^2$ harmonic
oscillators in the matrix basis, since $X_1$ and $X_2$ are
conjugated each other.

\section{Giant Graviton Fluid as Fractional Quantum Hall Fluid}

In this section we shall quantize the finite-matrix Hamiltonian
~(\ref{47d}) supplemented with Gauss law constraints derived
from from the Lagrangian ~(\ref{47}), to show that indeed the
1/2 BPS dynamics of giant gravitons in $AdS_5\times S^5$ allows
the formation of fractional quantum Hall (FQH) fluids with
filling factor $\nu=1/k$, with $k$ a positive odd number. There
are several approaches to quantize this model: (1) Canonical
approach with Gauss' constraint; (2) Path integral approach;
(3) Reduced canonical approach (first to solve the classical
constraints and then to quantize). Though they all lead to
essentially the same physics, each approach has its own
advantages in revealing some aspects of the underlying physics.

\subsection{``Classical'' quantum Hall droplet}

The Gauss law constraint derived from the Lagrangian~(\ref{47}),
\beq{56}
G=-i[X_1,X_2]+\Psi\Psi^\dag-\theta=
\frac{1}{2}[A,A^\dag]+\Psi\Psi^\dag-\theta=0,
\eeq
implies that the energy is discrete, even at the classical level,
with an energy gap of order $\theta$. To reveal this, we need to
solve the above constraint as well as the classical equations of
motion. Classically, inserting the solution of equation of motion,
\beq{56s}X_1+iX_2=A, \hspace{0.6in}
\Psi&=&e^{-i\mu_b t}\sqrt{M\theta}|v\rangle,
\eeq
with $A$ a constant $M\times M$ matrix and $|v\rangle$ a constant
vector of unit length, into the Hamiltonian~(\ref{47d}), we obtain
\beq{47b}H=\frac{B^2}{4m}{\rm Tr}A^\dag A+MB\theta \mu_b.
\eeq

In the oscillator basis, we may choose $|v\rangle=|M-1\rangle$.
The traceless part of the constraint~(\ref{56}) has many solutions
for $A$. Each of these solutions corresponds to a different
distribution of the giant gravitons on the 2-plane. For example,
the solution corresponding to the ground state is
\beq{56t}
A=\sqrt{2\theta}\sum_{n=1}^{M-1}\sqrt{n}|n-1\rangle\langle n|.
\eeq
It yields the radius squared matrix:
\beq{61} R^2&=&X_1^2+X_2^2=\frac{1}{2}(A^\dag
A+AA^\dag) \nn \\
&=&\sum_{n=0}^{M-2}\theta(2n+1)|n\rangle\langle n|
+\theta(M-1)|M-1\rangle\langle M-1|.
\eeq
Semi-classically the $M$ giant gravitons in this case are
uniformly distributed in the $(x_1,x_2)$ plane on a disk with
radius $\sqrt{2M\theta}$, (see Fig. 1a), with density
$\rho_{gg}=1/(2\pi\theta)$ and the inverse filling factor
$\nu^{-1}=B/(2\pi\rho_{gg})=B\theta$ in the analogy to the QH
``droplet''. Later we will show that this droplet can generally be
in a FQH liquid phase when the density is at the right value. The
total classical energy of this system is
$$E=\frac{B^2}{4m}\theta M(M-1)+MB\theta \mu_b.$$ The classical
energy contributed by the edge states is order $M\mu_b\sim 1/l_s$,
and can be ignored compared with the energy contributed by $X_i$.
This is consistent with the known fact that the edge excitations
are essentially gapless.
\begin{figure}[htpb]
  \centering
  \includegraphics{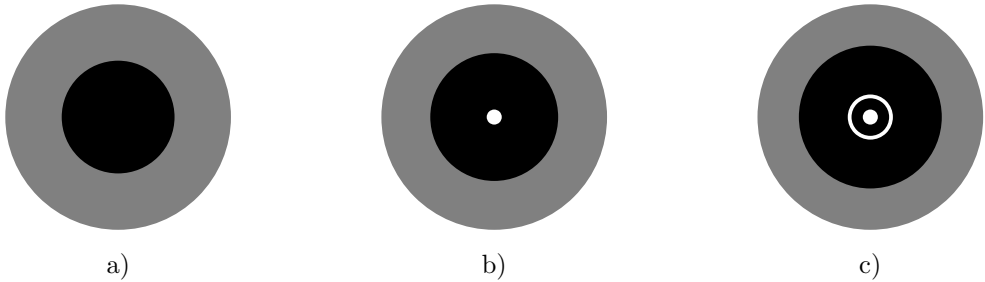}
\begin{minipage}{5in}
\caption{The giant gravitons in $AdS_5\times S^5$ and its analogy
to the QH ``droplet''. The gray region denotes the $AdS_5\times
S^5$ background. The giant gravitons are uniformly distributed on
the black region and the quasi-hole excitations in this giant
graviton background are denoted by the white region. a) The
(fractional) QH `droplet'' formed from the condensation of the
giant gravitons. b) A quasi-hole excitation at origin. c) Several
quasi-hole excitations.}
\end{minipage}
\end{figure}

Another interesting special solution is
\beq{60}
A=\sqrt{2\theta}\left(\sqrt{q}|M-1\rangle\langle 0|
+\sum_{n=1}^{M-1}\sqrt{n+q}|n-1\rangle\langle n|\right),
\eeq
where $q> 0$. It is known to be a quasi-hole excitation with
charge $-q$ (a defect) at the origin in the FQH background (Fig.
1b). It increases the radius of the state $|m\rangle$ from
$\sqrt{(2m+1)\theta}$ to $\sqrt{(2m+2q+1)\theta}$. This
corresponds to the excitation of the giant graviton at the $j=0$
state (with $j$ the angular momentum in LLL) to the $j=1$ state,
while the giant graviton at the $j=\ell$ state is excited to the
$j=\ell+1$ state, etc. The total number of giant gravitons is
unchanged. Finally, a rather general solution can be constructed
as follows: \beq{60a} A=\sqrt{2\theta}\sum_{i=1}^{m}
\left(\sqrt{q_i} |n_i\rangle\langle n_{i-1}|
+\sum_{n=n_{i-1}+1}^{n_i}\sqrt{n+q_i} |n-1\rangle\langle
n|\right), \eeq where
$|n_0\rangle=|0\rangle,\;|n_m\rangle=|M-1\rangle$ and $m\ll M$.
This solution can interpreted as several quasi-hole excitations
with charges $-q_i$, at different radii in the QH background
(Fig.1c). Each of these solutions corresponds to a classical,
stable configuration of the giant gravitons.

Finally, we remark that both the classical Hamiltonian and the
angular momentum depend on the trace of the radius-squared matrix
only. Therefore, both the energy and angular momentum are
quantized, even at the classical level.

\subsection{Gauss' law in the canonical approach}

To quantize the matrix model (\ref{47}), we treat the matrix
elements of $X_1$ and $X_2$ as operators, and impose the canonical
commutation relations \beq{48} [({\bf X}_1)_{mn}, ({\bf X}_2)_{rs}
]= \frac{i}{B} \delta_{mr}\delta_{ns}, \eeq because the Lagrangian
is first order in time derivative. Hereafter we shall use the
boldface letters to denote operators in the quantum theory. The
Hamiltonian for the ${\bf X}$'s is so ordered as to be given by
(${\bf Z}={\bf X}_1+i{\bf X}_2$)
\beq{51a} {\bf H}_X =
\frac{B^2}{4m} \sum_{mn}({\bf Z}_{mn}^\dagger {\bf
Z}_{mn}+\frac{1}{B}),
\eeq
corresponding to $M^2$ harmonic oscillators. The components of the
column vector ${\bf \Psi}$ correspond to $M$ oscillators and are
quantized to be bosons with the commutation relations
\beq{51} [{\bf \Psi}_m, {\bf \Psi}_n^\dagger]
=\frac{1}{B} \delta_{mn}.
\eeq
Further the operator ordering ambiguity in the quantum version of the
classical Gauss' law constraint~(\ref{56}) is fixed by requiring
that as quantum generators of unitary rotations of both ${\bf
X}_1$, ${\bf X}_2$ and ${\bf \Psi}$, the operator ${\bf G}$ should
satisfy the commutation relations of the $U(M)$ algebra.
Therefore, the traceless ($SU(M)$) part of the Gauss' law operator
${\bf G}$ can be constructed as the sum of two terms, ${\bf G}_X$
and ${\bf G}_{\Psi}$, which are the well-known bilinear
realization of the $SU(M)$ algebra in the Fock basis of bosonic
oscillators in the adjoint and fundamental representations,
respectively. \beq{53}
{\bf G} &\equiv& {\bf G}_X+{\bf G}_{\Psi}\nn \\
&=& -i\Bigl({\bf Z}_{mk}^\dagger {\bf Z}_{nk} - {\bf
Z}_{nk}^\dagger {\bf Z}_{mk}\Bigr)E_{mn} +{\bf \Psi}^\dagger_m
T^a_{mn}{\bf \Psi}_n T^a \eeq where $E_{mn}$ is the matrix with
only the element at the $m$-th row and $n$-th column being one and
all other elements zero; $T^a$ is the matrix for the generator of
$SU(M)$ in the fundamental representation. Then the traceless part
of the quantum Gauss's law is given by \beq{54} ({\bf G}^a_X+{\bf
G}^a_\Psi)|phys>=0. \eeq The $U(1)$ part of the quantum Gauss's
law, similar to that in quantum electrodynamics, just requires the
total $U(1)$ charge of the model to vanish:
\beq{55}
({\bf\Psi}_m^\dagger{\bf\Psi}_m-M\theta)|phys>=0.
\eeq
The two constraints~(\ref{54}) and (\ref{55}) together require
the physical states to be $U(M)$ singlets.

\subsection{Quantization of the inverse filling factor}

Because of purely group-theoretical reasons, Gauss' law
constraints (\ref{54}) and (\ref{55}) impose a severe restriction
on the possible value for the parameter $B\theta$, which is
closely related to the inverse filling factor $\nu^{-1}$ for giant
gravitons. Eq.~(\ref{54}) requires that the physical states are
invariant under $SU(M)$. Consequently, the representation of ${\bf
G}_X$ and that of ${\bf G}_\Psi$ must be conjugate to each other,
so that it is possible to form an $SU(M)$ singlet. The $X$-part is
known to be formed by the tensor product of the adjoint
representations, so it contains only the irreducible
representations with Young tableau having an integral multiple of
$M$ boxes. The same must be true for the conjugate representations
for the $\Psi$-part. Moreover, since eq.~(\ref{55}) is realized on
bosonic oscillators, it contains all the symmetric irreducible
representations of $SU(M)$, whose Young tableau consists of a
single row. Therefore, the number of boxes equals the eigenvalue
of the total number operator for $\Psi$-oscillators. In this
way~\cite{Poly01}, the Gauss' law plus group theory require that
$MB\theta$ be an integral multiple of $M$, or simply
\beq{56a}
B\theta=\ell=0,1,2,....
\eeq

This quantization in the Chern-Simons finite-matrix model can also
be viewed from several different angles. For example, it can be
understood as anomaly cancellation {\it a la} Callan and Harvey
\cite{Callan}. The anomaly here is referred as the Gauss law
anomaly, i.e. $[X_1,X_2]=i\theta$, can not be satisfied by any
finite matrices. Then one introduces the boundary degrees of
freedom and the anomaly is cancelled or ``compensated'' by the
boundary contributions when $B\theta$ is quantized. On the other
hand, one may also consider this as the finite-matrix model
version of the level quantization for the non-commutative
Chern-Simons (NCCS) term. This is because the second term of the
Lagrangian (\ref{47}) is of the form for the NCCS density:
\beq{57a}
\mathcal{L}= \frac{4\pi \kappa}{3} \theta
\int\; \epsilon_{\mu\nu\rho} {\rm Tr} D_\mu D_\nu D_\rho
\eeq
with the coefficient $\kappa=B\theta/4\pi$. Here we have viewed
the coordinates $X_i$ as the covariant derivative operators:
$X_i\sim \theta D_i$ and $D_0=-i\pa_t +A_0$, a trick often
used in the matrix model of D0-branes \cite{BFSS}. The level
quantization~\cite{ChenWu00,Level01} requires the level
$4\pi\kappa=\ell$ be integers, which is nothing but
eq.~(\ref{57a}). Finally, the same quantization has a
topological origin in path integral, by requiring the
complex density $\exp\{i\int dt L(t)\}$ in path integral
measure to be invariant under large $U(M)$ Gauge
transformations along a path with nonzero winding in the
compactified temporal direction. The 1d Chern-Simons term
${\rm tr} A_0$ changes under large $U(1)$ transformations,
so its invariance leads to the quantization (\ref{56a}).

The quantization condition~(\ref{56a}) immediately indicates
the quantization of the inverse filling factor, $\nu^{-1}$.
Classically $\nu^{-1}=B\theta$. However, there is a quantum
correction to this identification: namely, quantum
mechanically we have
\beq{57b}
\nu^{-1}=k=\ell+1=1,2,3,...\quad .
\eeq
Certainly this is consistent with the known fact that there is a
level shift, $\ell \to \ell+1$, in quantum U(1) NCCS
theory~\cite{ChenWu00,Level01}. For our finite-matrix model, the
quantum correction is due to the operator reordering effects in
the radius squared matrix, $R^2$, which semiclassically determines
the ``area'' occupied by the giant gravitons. Namely upon
quantization, the expression of Tr$R^2$ acquires an additional
``zero-point energy'' term, $M^2/B$, according to the quantization
condition ~(\ref{48}). This additional contribution to Tr$R^2$
increases the radius of each ``semi-classical orbit'' from the
classical value $\sqrt{(2n-1)\theta},\;(n=1,2,...,M)$ to
$\sqrt{(2n-1)(\theta+1/B)}$. Thus, the area of the giant graviton
droplet is increased due to quantum effects: ${\rm Area}\simeq
2\pi M(\theta+1/B)$ for $M\gg 1$. Since the filling factor is
defined by $\nu=2\pi M/(B\times {\rm Area})$, we obtain
$\nu^{-1}=\ell+1\equiv k$ by using the quantization
condition~(\ref{56a}).

So much for the mathematical derivation of the quantization
condition (\ref{57b}). Here some remarks on its physical meaning
are in order. The quantization condition (\ref{57b}) for the
filling factor $\nu$, on one hand, is similar in nature to the
Dirac quantization of the monopole charge, in that the allowed
values are special, directly related to the consistency of the
quantum dynamics. On the other hand, it is different from the
Dirac condition in that it deals with the behavior of a quantum
{\it many-body} system as a whole. Eq.~(\ref{4}) together with
(\ref{11}) tell us that $Br_m^2/2=\td{n}={\rm integer}$, where
$\td{n}$ is the number of the background fluxes in the area with
radius $r_m$. Then using previous discussions one has
\beq{63}\td{n}=B r_{_M}^2/2=kM.
\eeq
This result together eq.~(\ref{4}) yield the density of the giant
gravitons on $(x_1,x_2)$-plane:
\beq{58a}\rho_{gg}=\frac{1}{2\pi k}.
\eeq
For $k=1$, the giant graviton density is the same as
that of LLM fermions in the original IQH droplet for
$AdS_5\times S^5$. However, for $k>1$ the quantization
condition (\ref{57b}) tells us that something special
happens at the particular values (\ref{58a}) of the
giant graviton fluid density. A monopole charge that
does not satisfy the Dirac quantization condition
simply can not exist. However, a system of giant
gravitons with a density which does not satisfy the
quantization condition (\ref{57b}) still exists, but
they can not be in the particular states described
by the ground state of the Chern-Simons matrix model.
Namely self organization of giant gravitons into a
new incompressible quantum fluid may happen only at
the special densities given by eq. (\ref{58a}).

The above condition indicates that $\theta\sim l_s^2$.
No further condition exists to require $\nu$ must be
equal to 1, so in general the filling factor is
allowed to be a fraction $\nu=1/k$, with $k$ an integer.
Hence giant gravitons probing $AdS_5\times S^5$ can be
in a QH state with either integer or fractional filling
factor, depending on the density of giant gravitons on
the $(x_1,x_2)$ plane at $y=0$.

\subsection{Interactions between giant gravitons}

To determine the interaction between giant gravitons, we had
better adopt the reduced canonical approach. At the classical
level, the Gauss law constraint~(\ref{56}) can be solved
\cite{Poly01} in the eigenvalue basis of $X_1$ by
\beq{57}
\Psi&=&e^{-i\mu_b t}\sqrt{M\theta}|v\rangle,\hspace{0.5in}
|v\rangle=\frac{1}{\sqrt{M}}(1,1,...,1)^T, \nn \\
(X_1)_{mn}&=&x_n\delta_{mn},\hspace{0.5in}
(X_2)_{mn}=y_n\delta_{mn} +\frac{i\theta}{x_m-x_n}(1-\delta_{mn}).
\eeq
This solution implies that $x_m\neq x_n$ for $m\neq n$. It is
consistent with our previous argument that the world-volume gauge
symmetry of the giant gravitons breaks from $U(M)$ to $U(1)^M$.

Substituting the solution (\ref{57}) into the classical
Hamiltonian (\ref{47b}), one obtains the classical
Hamiltonian in terms of the variables $x_m$ \cite{Poly01}:
\beq{58}
H_{ca}=\frac{1}{m}\sum_{m=1}^M (p_m^2+\frac{B^2}{4}x_m^2)
+\frac{1}{4m}\sum_{n\neq m} \frac{(B\theta)^2}{(x_m-x_n)^2}.
\eeq
$H_{ca}$ is nothing but the integrable one-dimensional Calogero
model \cite{Cal69} for non-relativistic particles on a line.

Then we want to quantize the system. The correct way is not to
impose the canonical commutation relations directly to the
variables $x_n$ and $p_n$ in eq.~(\ref{58}). We should solve the
quantum version of Gauss law constraint in the eigenvalue basis of
${\bf X}_1$, and then substitute the solution into the
Hamiltonian~(\ref{47b}). Non-commutativity of the quantum
operators shifts the strength of the two-body inverse square
potential from $\ell^2$ to $\ell(\ell+1)$ \cite{Poly01}, or
in terms of the inverse filling factor $k$:
\beq{59}
{\bf H}_{ca}=\frac{1}{m}\sum_{m=1}^M (-\frac{\pa^2} {\pa
x_m^2}+\frac{B^2}{4}x_m^2) +\frac{1}{4m}\sum_{n\neq m}
\frac{k(k-1)}{(x_m-x_n)^2}.
\eeq
The shift in the strength of the two-body inverse square
potential is apparently related to the level shift,
$\ell \to \ell+1$, in U(1) non-commutative Chern-Simons
theory \cite{ChenWu00,Level01}.

It is well-known that with $k=1$ this model describes a free
fermion system \cite{Suth71}, which corresponds to the $\nu=1$
IQH droplet in phase space and agrees with the observations
made in the LLM geometry\cite{LLM} and in the matrix model
for AdS/CFT\cite{Ber04a} for 1/2 BPS geometries. The essential
information we obtain from the Hamiltonian (\ref{58}) in
the present context is that the two-body interactions
between giant gravitons due to quantum fluctuations of
stretched open strings are {\it repulsive} for $k>1$. In
condensed matter physics, the repulsive nature of the
interactions is considered to be a crucial condition for
the formation of new incompressible QH fluid states. Indeed,
the ground-state wave function of the Calogero
model
\beq{Ca}
\Psi_0(x_1,x_2,\cdots) = \prod_{m<n} (x_m-x_n)^k \exp
\{-\frac{B}{2} \sum_n x_n^2 \},
\eeq
is nothing but the 1d representation (in the Landau gauge) of
the Laughlin wave function in the LLL with $\nu=1/k$ on a disk
geometry. Similar correspondence exists for excited states.
(For more discussions on the explicit relationship between the
Calogero model and the FQHE, see ref. \cite{Azuma93}.) Thus,
from the quantum Calogero model obtained in our present context,
we have demonstrated that the ground state of the QH system of
giant gravitons at the right value of density, $\nu=1/k$ with
$k>1$ integer, is in a strongly correlated quantum fluid phase,
an incompressible FQH fluid of giant gravitons! (Actually $k$
should be an odd integer; see next subsection.)

\subsection{New Area Quantization and Stringy Exclusion Principle}

To gain more insight into the FQH state of giant gravitons with
$k>1$, we note that
\beq{62} \langle n|R^2|n\rangle
-\langle n-1|R^2|n-1\rangle =2\theta. \eeq
Thus a giant graviton in the FQH droplet with $k>1$ occupies
a $k$ times bigger area than the $k=1$ case. This
predicts a new area quantization for the FQH state of
giant gravitons. Previously in the fluid description we have
already seen that the density of giant gravitons in the FQH
droplet is $k$ times bigger than that of LLM fermions in the
original IQH droplet. But that is only a coarse-grained
description on average. The result (\ref{62}) in the matrix model
is a property at the microscopic level. This microscopic property
is a compelling evidence that the ground state of the Chern-Simons
matrix model (\ref{47}) corresponds to a new incompressible
quantum fluid state, a FQH fluid state. This makes the FQH state
distinct from other possible states of the system with the same
fractional filling factor, say from a dilute gas phase which is
compressible. We expect that the stability of the FQH state may
survive from the back-reactions of the giant gravitons. Namely the
new area quantization (\ref{62}), which is $k$ times bigger than
that in the smooth 1/2 BPS LLM's geometries, should emerge in the
new geometry that arises due to the back-reactions of giant
gravitons.

This is a new aspect of the giant graviton physics. It implies a
{\it stringy exclusion principle} at work. This stringy exclusion
principle differs from the one discussed previously
\cite{MaldStro,MST00}, in that it deals with a strong correlation
effect of many giant gravitons, rather than a property of a single
giant graviton. This stringy exclusion principle is actually an
analogy of the generalized exclusion principle in condensed matter
physics \cite{Hald91,Wu94} that is at work in the FQH state.

We may also examine the state (\ref{60}) with quasi-hole
excitations. The value of the quasi-hole charge $-q$ is arbitrary
in the classical theory. It can be shown \cite{Poly01} that, to
accompany the quantization of the filling factor $\nu$, there is
also a quantization of $q$, with the minimal value $q=1/k$. By the
standard treatment of the Calogero model (\ref{59}), we expect
that in addition to the fractional charge $q=1/k$, the quasi-holes
in the FQH liquid state of giant gravitons should have both
fractional exchange (anyon) statistics with $\theta_{stat}=\pi/k$
\cite{Poly89,Iso}, and fractional exclusion statistics with
$\lambda_{qh}=1/k$. This value of exclusion statistics parameter
for quasi-holes is just dual to the exclusion statistics parameter
of the giant gravitons in the FQH state, $\lambda_{gg}=k$
\cite{BerWu,WuYu}.

The possible values of the inverse filling factor $k$ can be
further constrained by consideration of quantum statistics.
From the exchange statistics in one dimension in the sense
of ref. \cite{Poly89} in terms of the scattering phase shift,
$\exp (-i\pi k)$, $k$ should be odd because the constituent
LLM particles are fermions. The same restriction can be
obtained by considering anyon statistics for quasi-hole in
two dimensions, which is the same as that in the Calogero model
\cite{Azuma93}. The argument goes as follows \cite{TaoWu}:
Consider a cluster of coincident $k$ quasi-holes. It has a charge
just opposite to that of the constituent giant graviton, so it is
the same as the charge of the original LLM particle. If in the
spectrum the only states that carry the same charge are those of
the hole of constituent giant gravitons, or the LLM particle, the
cluster of coinciding quasi-holes should be identified to be an
original LLM particle, including their statistical behavior. The
statistics of a cluster of $k$ anyons is known \cite{Wu84} to be
$k^2\theta_{stat}$, which is just $k\pi$. It should be the same as
that of the original LLM particle which is known to be a fermion.
So we should choose
\beq{odd}
k=odd=1,3,5,\cdots
\eeq
for the giant graviton FQH fluids.

\subsection{Back-reaction of giant gravitons: speculations}

In above study of 1/2 BPS dynamics of giant gravitons in the
$AdS_5\times S^5$ background, the back-reaction of giant gravitons
is neglected. This approximation is valid if the number $M$ of the
giant gravitons satisfy $M\ll \td{N}$; $\td{N}$ is the number of
the R-R 5-flux in $AdS_5\times S^5$ background. This implies that
the size of the giant graviton droplet, of the order of
$\sqrt{g_sM} l_s$ when all fundamental constants are restored, is
much smaller than that of the IQH droplet, $\sqrt{g_s\td{N}} l_s$.
Due to the extraordinary stability of the FQH state, one expects
that the emergence of the FQH state and many of their properties,
such as new area quantization and stringy exclusion principle for
quasi-holes etc, may survive from the back-reactions of giant
gravitons. In this subsection we speculate on this possibility.

In subsection V.A and V.C, we have seen that $M$ giant gravitons
are distributed semi-classically in a disk on the $(x_1,x_2)$
plane with uniform density $1/(2\pi k)$. This semi-classical
result is expected to survive in the full quantum theory. When
$k=1$, this QH droplet vacuum corresponds to a quasi-hole disk as
a $z=1/2$ region in the center of the $\nu=1$ IQH droplet in the
LLM geometry \cite{LLM}. This recovers the known result that the
back-reactions of giant gravitons in the IQH state with $k=1$ give
rise to new smooth 1/2 BPS geometries \cite{LLM}.

For $k>1$ case, the droplet of giant gravitons has a $k$ times
larger area than the $k=1$ case in the center of the LLM's IQH
background on the $(x_1,x_2)$ plane. Accordingly, the droplet of
giant gravitons has a smaller density so it is a {\it fractional}
quantum Hall fluid with giant graviton filling factor
$\nu_{gg}=1/k$. In the region of the giant graviton FQH droplet,
we have actually a two-component fluid, with the second component
being the original IQH background fluid, so that the total density
of the original fermions in the region of the FQH droplet of the
giant gravitons is $\rho=(k-1)/(2\pi k)$. Suppose that the
back-reaction of the giant gravitons in such a FQH state can be,
at least in a certain decoupling limit, restricted to the 1/2 BPS
sector of geometries then a new 1/2 BPS geometry in IIB
supergravity will appear due to the back-reactions. Since the FQH
state with $k>1$ corresponds to a QH droplet with a region in
which the density of original fermions is between 0 and 1, the
corresponding LLM solutions will have a null singularity,
according to refs. \cite{LLM,BSY05,Loughlin05}. This null
singularity is believed to be resolved by possible local quantum
effects~\cite{MT01,HS05,BJS05} and has well-defined description in
dual CFT\cite{HHI00}. If the back-reactions in the full IIB
supergravity can not be restricted semi-classically to the 1/2 BPS
sector, then orbifold singularities are possible to arise in the
less-than-half BPS cases. So in general we expect that
back-reactions of giant gravitons in a FQH state would lead to the
emergence of geometries with (naked) singularities.

Since a FQH state is a well-behaved quantum state, which maintains
long-distance quantum coherence (more precisely, algebraic
off-diagonal long-range order) \cite{Girvin87,Hald88}, we believe
quantum effects in IIB string theory on the corresponding geometry
should resolve the singularity. The properties of the FQH state
may help us understand the underlying mechanism in quantum gravity
for the resolution of singularity.

\section{Summary}

In this paper, we re-examined the dynamics of giant gravitons
probing $AdS_5\times S^5$ geometry, inspired by LLM's 1/2 BPS
geometry in IIB supergravity. The giant gravitons we examined are
D3-branes wrapping on a three-sphere inside $S^5$ and, in the
spirit of mini-superspace approximation for the moduli space of
the 1/2 BPS geometries, their dynamics is restricted to that on a
two dimensional plane in $AdS_5\times S^5$. The non-abelian
low-energy dynamics of the giant graviton probes is shown to lead
to a Chern-Simons matrix model, which describes the fluid dynamics
of a QH system, i.e. a system of non-relativistic charged
particles moving in a strong constant magnetic field. In terms of
incompressible giant graviton fluid, the gauge symmetry of the
matrix model is broken from $U(M)$ down to $U(1)^M$, and an
additional non-commutative constraint is supplemented to the
matrix model action.
The D-brane nature of the giant gravitons dictates that edge
states are excited when the boundary of the giant graviton droplet
fluctuates. Indeed, the low energy world-volume theory for a giant
graviton fluid reduces to a Chern-Simons finite-matrix model,
which is known to incorporate the edge excitations of a QH droplet
of finite size, with either the integer or the fractional filling
factor. We discussed various physical analogies between the giant
graviton fluid and the FQH liquid, and speculated the possibility
that the FQH nature of the giant graviton fluid may survive beyond
the probe approximation and give rise to new geometry with
singularity in IIB supergravity. In short, our results demonstrate
that the giant gravitons, as quasi-hole excitations in
$AdS_5\times S^5$ droplet background, can condense into a new
quantum Hall fluid with filling factor $\nu=1/k$ for $k$ an odd
integer.

It is interesting to ask whether there is a QH analogy in the open
string fluid (mentioned in sect. IV.A), in which the $U(M)$ gauge
symmetry is unbroken. The experience with the SYM matrix
model\cite{Ber04a} indicates that the answer should be ``YES'',
because there is a one-to-one correspondence between the matrix
basis ($M^2$ uncoupled bosonic oscillators) and the eigenvalue
basis ($M$ free fermions) in the matrix model. To show the QH
analogy explicitly, we note that each of $M^2$ uncoupled identical
oscillators behaves exactly the same as a charged particle in a
magnetic field. Then we may introduce the non-commutativity
constraint again by assuming the system is in an incompressible
fluid state. (This assumption is self-consistent if the
interactions are repulsive.) The bosonic statistics will {\it not}
interfere with the formation of an incompressible fluid provided
interactions are repulsive, which make the multi-occupied single
particle states unfavorable energetically. So the system can be in
an incompressible quantum fluid, analogous to bosonic Laughlin
states with filling fraction $\nu=1/k$ for even $k$.

One may also consider D3-brane probes wrapping on $S^3$ inside
$AdS_5$. These giant gravitons are located in the exterior of the
circle $r^2=r_0^2$, and are viewed as quasi-particle excitations
of the $AdS_5\times S^5$ droplet. As charged particles, they also
couple to a constant background magnetic field, but now the number
of the magnetic flux is $N$ instead of $\td{N}$ (see (\ref{4}) and
(\ref{5})). This is a system dual to the one considered in this
paper. It would be interesting to study the condensation of these
giant gravitons and ask whether there is a QH analogy for this
system.

We conclude this paper with a short comment on the emergent QHE
from the $\CN{4}$ SYM side in AdS/CFT correspondence.
Ref.~\cite{GMSS05} made a remarkable inspiring attempt in this
direction. The 1/2 BPS sector for a single chiral scalar in
$\CN{4}$ SYM compactified on $S^3$ is a matrix model. A
Chern-Simons term emerges in this matrix model if an internal
circle is boosted in the bulk theory as that shown by the last
equation in (\ref{9}). So it is an educated guess that, following
the procedures presented in this paper, we will end up with FQH
systems with filling fractions of the form $1/k$ for odd $k$
\cite{DWW}.

\acknowledgments{The authors thank Li-Sheng Tseng and Eric Sharpe
for helpful discussions. X.-J. Wang was partly supported by China
NSF, Grant No. 10305017, and through USTC ICTS by grants from the
Chinese Academy of Science and a grant from NSFC of China. Y.-S.
Wu was supported in part by the US NSF through grant PHY-0457018.
He also thanks the Institute for Solid State Physics, University
of Tokyo, Prof. Ueda and Prof. Kohmoto for the warm hospitality
that he received during his sabbatical visiting.}

\end{document}